\documentclass[twocolumn,aps,prd,showpacs,epsfig,nofootinbib]{revtex4-1}
\usepackage{graphicx,epstopdf,latexsym,amssymb,amsmath,color,mathrsfs,rotating,multirow}
\usepackage[center]{subfigure}
 \usepackage[colorlinks,linkcolor=blue,citecolor=blue,urlcolor=blue]{hyperref}

\begin{document}
\allowdisplaybreaks
 \newcommand{\bq}{\begin{equation}}
 \newcommand{\eq}{\end{equation}}
 \newcommand{\bqn}{\begin{eqnarray}}
 \newcommand{\eqn}{\end{eqnarray}}
 \newcommand{\nb}{\nonumber}
 \newcommand{\lb}{\label}
 \newcommand{\f}{\frac}
 \newcommand{\p}{\partial}
\newcommand{\PRL}{Phys. Rev. Lett.}
\newcommand{\PLB}{Phys. Lett. B}
\newcommand{\PRD}{Phys. Rev. D}
\newcommand{\CQG}{Class. Quantum Grav.}
\newcommand{\JCAP}{J. Cosmol. Astropart. Phys.}
\newcommand{\JHEP}{J. High. Energy. Phys.}

\title{Universal features of quantum bounce in loop quantum cosmology}

\author{Tao Zhu${}^{a, b}$}
\email{Tao$\_$Zhu@baylor.edu}

\author{Anzhong Wang${}^{a, b}$\footnote{The Corresponding Author.}}
\email{anzhong$\_$wang@baylor.edu}

\author{Klaus Kirsten${}^{c}$}
\email{klaus$\_$kirsten@baylor.edu}

\author{Gerald Cleaver${}^{d}$}
\email{gerald$\_$cleaver@baylor.edu}

\author{Qin Sheng${}^{c}$}
\email{qin$\_$sheng@baylor.edu}

\affiliation{${}^{a}$ Institute for Advanced Physics $\&$ Mathematics, Zhejiang University of Technology, Hangzhou, 310032, China\\
${}^{b}$ GCAP-CASPER, Physics Department, Baylor University, Waco, TX 76798-7316, USA\\
${}^{c}$ GCAP-CASPER, Mathematics Department, Baylor University, Waco, TX 76798-7328, USA\\
${}^{d}$ EUCOS-CASPER, Physics Department, Baylor University, Waco, TX 76798-7316, USA}

\date{\today}

\begin{abstract}

In this Letter, we study analytically  the evolutions of the  flat Friedmann-Lemaitre-Robertson-Walker (FLRW) universe and its linear  perturbations 
in the framework of  {\em the  dressed metric approach} in loop quantum cosmology (LQC). Assuming that the evolution of the background is 
dominated by the kinetic energy of the inflaton at the quantum bounce, we find that both evolutions of the background and its perturbations  
are independent of the inflationary potentials during the pre-inflationary phase. During this period the  effective potentials of the perturbations  can 
be well approximated by a P\"oschl-Teller (PT) potential, from which we find analytically the mode functions and then calculate the corresponding 
Bogoliubov coefficients at the onset of the  slow-roll inflation, valid for any inflationary model with a single scalar field.  Imposing the Bunch-Davies 
(BD) vacuum in the contracting phase  prior to the bounce when the modes are all inside the Hubble horizon, we show that particles are generically  
created due to the pre-inflation dynamics.  Matching them to those obtained  in the slow-roll inflationary phase, we investigate the effects of the  
pre-inflation dynamics on the scalar and tensor power spectra and find features that can be tested by current and forthcoming observations. In 
particular, to be consistent with the Planck 2015  data, we find that the universe must have expanded at least $141$ e-folds since the bounce.

\end{abstract}


\maketitle

\section{Introduction }

The paradigm of cosmic inflation has achieved remarkable successes in solving several problems of the standard big bang cosmology and predicting 
the primordial perturbation  spectra whose evolutions explain both the formation of the large scale structure of the universe and the small inhomogeneities 
in the cosmic microwave background (CMB) \cite{inflation}. Now they are matched to observations with  unprecedented precisions \cite{WMAP, Planck2013, Planck2015}. 
However, such successes are contingent on the understanding of physics in much earlier epochs when energies were about the Planck scale. This leads to several conceptual issues. 
For example, to be consistent with observations,   the universe must have expanded  at least $60$ e-folds during its inflationary phase. 
  {However,  if the universe had expanded a little bit more than $70$ e-folds during inflation (as it is the case in a large class of inflationary models \cite{MRV}), 
then one can show that the wavelengths of all fluctuation modes which are currently inside the Hubble radius were smaller
than the Planck length at the beginning of the period of inflation. This was referred to as the trans-Planckian issue in  \cite{trans-planck}, and
 leads to the question about the validity of the assumption:  {\em the matter fields are quantum in nature but 
the spacetime is still  classical}, which are  used  at the beginning of inflation in order to make predictions \cite{inflation}. 
In addition, insisting on the use of general relativity (GR) to describe the inflationary process will inevitably lead to an initial singularity \cite{singularity}. 
Moreover, the inflation paradigm usually sets the BD vacuum state at the time when the wavelength of fluctuations were well within the Hubble 
horizon during the inflationary process. However, such treatment ignores the pre-inflationary dynamics which could lead to non-BD states at the onset of inflation, even when 
these modes were well inside the Hubble horizon during inflation. For more detail about the sensibility of the inflationary  paradigm to Planckian physics, we refer the readers to \cite{trans-planck,DB}. }

All the issues mentioned  above are closely related to the fact that we are working in the regime where GR is known to break down. One believes that new physics 
in this regime - a quantum theory of gravity, will provide a complete description of inflation as well as its pre-inflationary dynamics. LQC is one of such theories that 
offers a framework to address these issues, in which the inflationary scenarios can be extended  from the onset of the slow-roll inflation back 
to the Planck scale in a self-consistent way \cite{planck_extension,planck_extension_CQG,quadratic_loop}. Remarkably, the quantum geometry effects of LQC at the Planck scale provide a 
natural resolution of the  big bang singularity (see \cite{bounce,Ashtekar2015CQG,BB16, Yang_alternative_2009} and references therein). In such a picture, the singularity 
is replaced by a quantum bounce, and the universe that starts at the bounce can eventually evolve to the desired slow-roll inflation 
\cite{AS10,bounce_inflation, bounce_inflation2, deformed_tensor, deformed_scalar,deformed,Starobinsky_loop,bounce_effects}. An important question now is whether the 
quantum bounce can leave any observational signatures to current/forth-coming observations, so LQC can be placed directly under experimental tests. The answer to this question is 
affirmative.   { In fact, with some (reasonable) assumptions and choice of the initial conditions, the {\em deformed algebra approach} already leads to inconsistence 
with  current observations \cite{deformed}.} Note that in  general there are two main approaches  to implement cosmological perturbations in the framework of LQC, the 
{\em dressed metric} and {\em deformed algebra approaches} \cite{bounce,Ashtekar2015CQG,BB16}. In both, the primordial perturbations  have been intensively studied {\em numerically} 
\cite{planck_extension_CQG,quadratic_loop, bounce_effects, deformed, deformed_scalar, deformed_tensor, Starobinsky_loop}.

One of our purposes of this Letter,    in contrast to the previous numerical studies, is to present an {\em analytical} analysis of the effects of the quantum bounce and pre-inflation dynamics 
on the evolutions of both background and  spectra of the scalar and tensor perturbations, in the framework of  the  dressed metric approach \citep{planck_extension,planck_extension_CQG,quadratic_loop}. 
It is expected
that such an analysis will provide a more complete understanding of the problem and deeper insights. In the following, we will focus on the case that the kinetic energy of the inflaton
dominates the evolutions at the bounce, because a potential dominated bounce is either not able to produce the desired slow-roll inflation \cite{Starobinsky_loop},  or leads to a large amount of e-folds of expansion.
  { This will wash out all the observational information about the pre-inflation dynamics and the resulting perturbations are the same as those 
given in GR \citep{bounce, Ashtekar2015CQG,BB16}. Assuming that   the influence of the potential at the bounce is negligible, } our studies show that: 
{
 \begin{itemize}
 
 \item  { During the pre-inflationary phase, the evolutions of the background and the scalar and tensor perturbations are    independent of the inflationary
potentials.  Thus,   the  evolution of the background is the same for any chosen  potential, and in this sense we say that it is {\em universal}. }

\item  { During this phase the potentials of the scalar and tensor perturbations  can be well approximated by an effective PT potential, for which  analytic solutions of the mode functions can be found. 
 The Bogoliubov coefficients at the onset of the slow-roll inflation can thereby be calculated  [cf. (\ref{main})],  which are valid for any slow-roll inflationary model with a single scalar field. Assuming that the universe is in the BD vacuum in 
 the contracting phase (the moments where $t \lesssim - t_s$ as shown in Fig. \ref{length})  we find  that particle creations  occur generically during the pre-inflation phase. }
 
\item Oscillations  always happen in the power spectra, and their phases for both scalar and tensor perturbations are   the same, in contrast to other theories of quantum gravity  \cite{trans-planck,Zhu1}.

\item Fitting the power spectra to   the Planck 2015  data \cite{Planck2015}, we find the lower bound for $N_{\text{tot}} \equiv \ln{(a_0/a_B)} >141$ (95\% C.L.),
 where $a_B$ and $a_0$ denote the expansion factor at the bounce and current time, respectively. Details of the  calculations will be reported elsewhere \cite{bounce_uniform}.
\end{itemize}

 \section{Quantum Bounce}

In LQC, the semi-classical dynamics of a flat FLRW universe with a single scalar field $\phi$ and potential $V(\phi)$  is described by
\cite{planck_extension,planck_extension_CQG,quadratic_loop},
\bqn
\lb{friedmann}
&&H^2=\frac{8\pi}{3m_{\text{Pl}}^2}\rho\left(1-\frac{\rho}{\rho_\text{c}}\right), \\
\lb{klein}
&&\ddot \phi+3 H \dot \phi+V_{,\phi}=0,
\eqn
where $H\equiv \dot a/a$ is the Hubble parameter, 
a dot denotes  the derivative with respect to the cosmic time $t$, and $\rho_c$ is the maximum energy density, with $\rho\equiv \dot\phi^2/2+V(\phi) \le \rho_c$. Eq.~(\ref{friedmann}) shows
 that the big bang singularity  now is replaced by a non-singular quantum bounce at $\rho=\rho_\text{c}$ [cf. Fig.~\ref{scalar_factor}].
The background evolution has been extensively studied, and one of the main results is that, following the bounce, a desired slow-roll inflation phase is
almost inevitable,  {provided that the evolution is dominated initially by the kinetic energy of the scalar field  at the quantum bounce \cite{bounce, bounce_inflation, bounce_inflation2, Starobinsky_loop}.
In this Letter, we will focus on this case. 
Then, ignoring the potential term $V(\phi)$, from Eqs.~(\ref{friedmann}) and (\ref{klein}) we find 
\bqn
\lb{Bsolution}
a(t)=a_\text{B}\left(1+\gamma_B \frac{t^2}{t_{\text{Pl}}^2}\right)^{1/6},\lb{ana_background}
\eqn
where $a_\text{B} \equiv a(t_{B}), \gamma_B\equiv {24\pi \rho_{\text{c}}}/{m_{\text{Pl}}^4}$,
and $t_{\text{Pl}}$ denotes  the Planck time. In writing the above expression we also set $t_{\text{B}}=0$.
In Fig.~\ref{scalar_factor} we display the above analytical solution and the equation of state 
\bq
\lb{EoS}
w_{\phi} \equiv \frac{\dot{\phi}^2 - 2V(\phi)}{\dot{\phi}^2 + 2V(\phi)},
\eq
together with several numerical solutions of $a(t)$ for different potentials.  From this figure, specially the curves of $w_{\phi}$, 
 we can see that the universe  experiences three different phases:  {\em bouncing,  transition, and slow-roll inflation}. During the bouncing phase, 
 $w_{\phi}$ remains almost one until $t/t_{Pl} \simeq 10^4$. Then, it suddenly drops from 1 to -1 at $t/t_{Pl} \simeq 10^5$.
 This transition phase is very short in comparison to the other two, and the kinetic energy of the scalar field drops almost 12 orders from the beginning of this phase to the 
 end of it. Afterwards, the  potential energy $V(\phi)$ dominates the evolution, and  $w_{\phi}$ remains 
 practically  $-1$ during the whole slow-roll inflation phase. The end of this transition phase   can be  well defined as the moment where
 $\ddot{a}(t = t_i) = 0$, as shown in Fig. \ref{scalar_factor}. Afterward, the expansion of the universe will be accelerating $\ddot{a}(t > t_i) > 0$.
 However, unlike   $t_i$, the starting point of the transition phase is not abrupt, even though the division  is very clean in concept, as one can see from Fig. \ref{scalar_factor}. Fortunately, the results are not sensitive 
 to such a choice at all, as argued  below and shown in detail in \cite{bounce_uniform}.  In particular, we find that the choices 
of $w_{\phi} = 0.95$  and $w_{\phi} = 2/3$ make no (observational) difference in the power spectra and the total e-folds of the expansion of the universe. }

 During  the bouncing phase,  {\em the evolution of
 $a(t)$ is independent of the choice of  $\phi_\text{B}$ and  the choice of  the  potential $V(\phi)$ of the scalar field. } 
This is  because $V(\phi)$ remains very small and the kinetic energy is completely dominant during this whole phase. For example, for the potential $V(\phi) = V_0\phi^n$ with $n=2$, we find that
$V(\phi)/m_\text{Pl}^4 \in(2\times 10^{-11}, 4.5\times 10^{-11})$; for $n=1/3$,  $V(\phi)/m_\text{Pl}^4 \in(9 \times 10^{-12},
1.2\times 10^{-11})$; and for the Starobinsky potential,  we have $V(\phi)/m_\text{Pl}^4 \in (7\times 10^{-13}, 7.3\times 10^{-13})$. 
This explains why the evolution  of $a(t)$  is universal during this period.

\begin{figure}
\includegraphics[width=8.5cm]{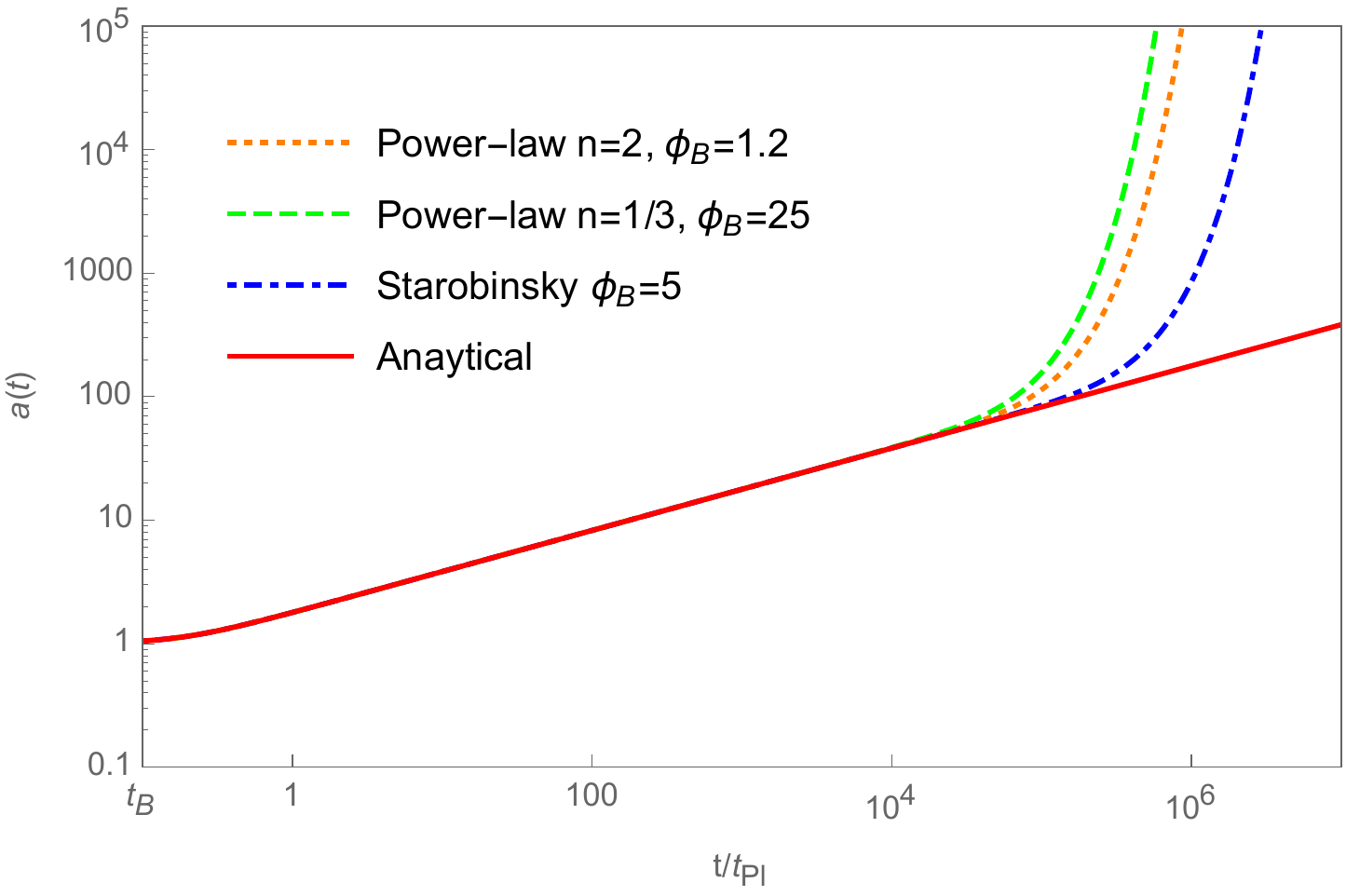}\\
\includegraphics[width=8.5cm]{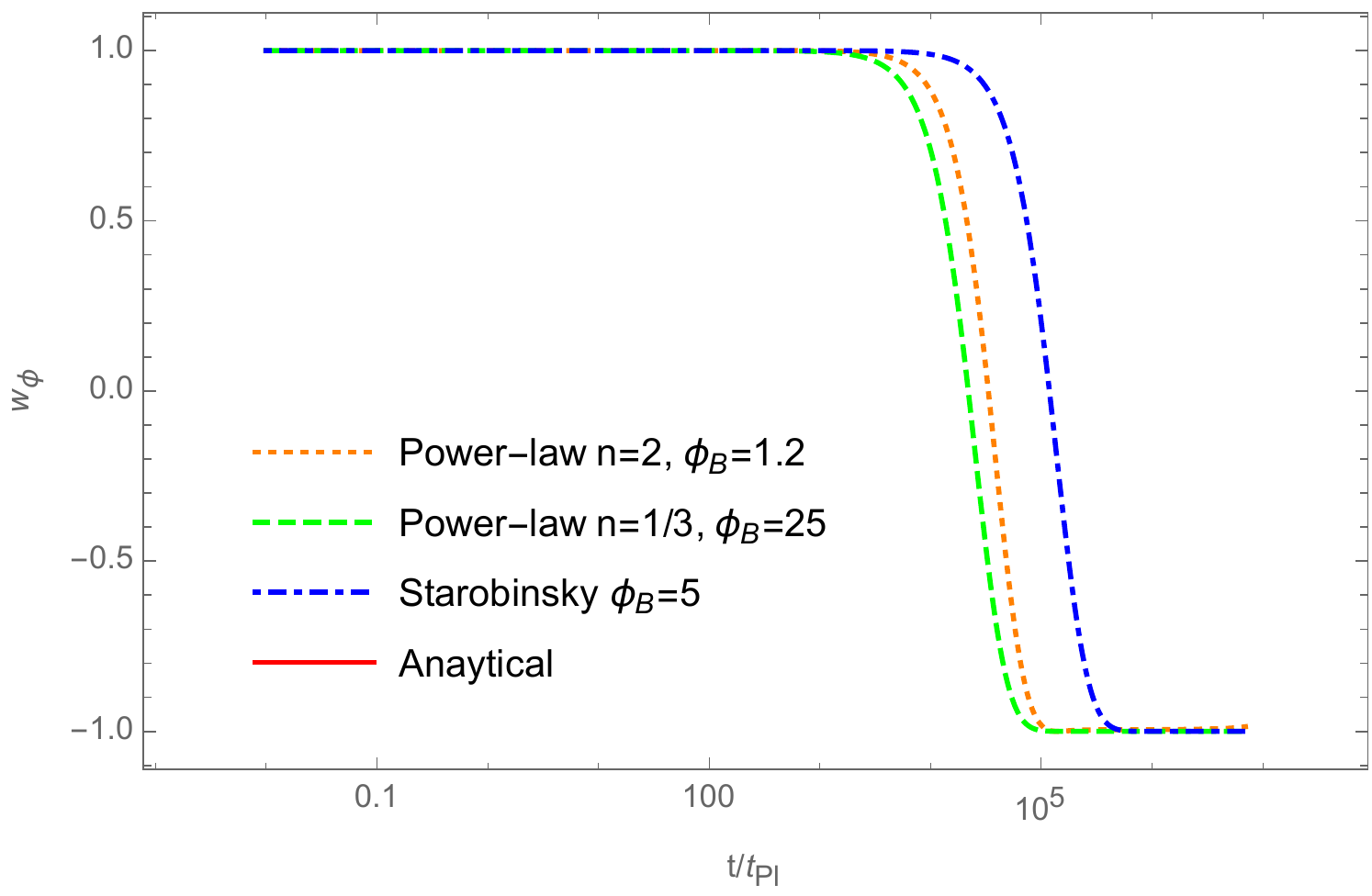}
\caption{Evolutions of $a(t)$  and $w_{\phi}$ for the power-law $V(\phi)=\frac{1}{2}m^{4-n} \phi^{n}$ and Starobinsky  $V(\phi)=\frac{3}{4} M^2 M_{\text{Pl}}^2 (1-e^{-\sqrt{2/3} \phi/M_{\text{Pl}}})^2$ potentials.
Solution (\ref{ana_background}) is also shown.
We choose $ m=1.3\times 10^{-6}$ for $n=2$, $m=1.1\times 10^{-3}$ for $n=1/3$, and $M=2.5\times 10^{-6}$ for the Starobinsky potential. In all the cases we set $m_\text{Pl} =1$.} \label{scalar_factor}
\end{figure}

 \section{  Primordial Power Spectra}

  The linear perturbations in
the  dressed metric approach \cite{planck_extension,planck_extension_CQG} were studied numerically in detail  with the inflationary potential
$V(\phi) = m^2\phi^2/2$ \cite{quadratic_loop}. In this Letter,  {our goals   are two-fold: First,  we study these perturbations {\em analytically}, and provide their explicit expressions.
Second, we show that they are  independent of the choices of the slow-roll inflationary potentials, so they are {\em universal}. In fact, 
this follows directly from the universality of the evolution of $a(t)$ during this phase.} To show this, let us start with the scalar and tensor perturbations
\cite{planck_extension, planck_extension_CQG, quadratic_loop},
\bqn\lb{scalar_eom}
\mu_k^{(s, t)}(\eta)''+\left(k^2-\frac{a''}{a}+ U^{(s, t)}(\eta)\right)\mu_k^{(s, t)}(\eta)=0,
\eqn
where
$U^{(s)} (\eta) \equiv a^2 \left(\mathfrak{f}^2 V(\phi)+2 \mathfrak{f} V_{,\phi}(\phi)+V_{,\phi\phi}(\phi)\right)$, $U^{(t)}(\eta) = 0$,
with $\mathfrak{f}\equiv \sqrt{24 \pi G}\dot \phi/\sqrt{\rho}$. $\mu_k^{(s, t)}(\eta)$ denote the Mukhanov-Sasaki variables with $\mu_k^{(s)}(\eta)=z_s \mathcal{R}$
and $\mu_k^{(t)}(\eta)=a h_k/2$, where $\mathcal{R}$ denotes the comoving curvature perturbations,   $h_k$ the tensor perturbations, and $z_s\equiv a \dot \phi/H$.
A prime denotes the derivative with respect to the conformal time $\eta(t)=\int_{t_{\text{end}}}^t  dt'/a(t')$,
where $t_{\text{end}}$ is the time when the inflation ends.  Near the bounce,
$U^{(s)}(\eta)$ is negligible \cite{planck_extension_CQG,quadratic_loop,bounce_uniform}. During the transition phase,
$\rho$ drops down to about $10^{-12} \rho_\text{c}$, and  $(a''/a-U^{(s)}(\eta))\to z_s''/z_s$,
so thereafter  the perturbations reduce precisely to those of GR \cite{planck_extension_CQG, Ashtekar2015CQG}.

The evolutions of the perturbations depend on both  background and   wavenumber $k$. As we consider only the case in which the kinetic energy dominates the evolution of
the background at the bounce, both scalar
and tensor perturbations follow the same equation of motion during the bouncing phase ($t/t_{Pl} \le 10^{4}$). In this case, the term $a''/a$ in Eq.~(\ref{scalar_eom})
defines a typical radius $\lambda=\sqrt{a/a''}$ for $a'' > 0$, which plays the same role as that of the comoving Hubble radius $L_H=(aH)^{-1}$ often used in GR.
However, for a better understanding, we find that here it is more proper to use  $a/a''$, as shown schematically in Fig.~\ref{length}. For example, when the modes
are inside the radius ($1/k^2< \lambda^2 $), the solution of Eq.~(\ref{scalar_eom}) is of the form,  $ e^{\pm i \int \sqrt{k^2-a''/a} d\eta}$. When the modes are
outside of the ``horizon" (radius) ($1/k^2> \lambda^2$), it is of the form,  $e^{\pm \int \sqrt{a''/a-k^2}d\eta }$. The term $a''/a$ has its maximum at the bounce,
$a''/a|_{t=0}= a_{\text{B}}^2 \gamma_B m_{\text{Pl}}^2/3$, which defines a typical scale $k_\text{B}=\sqrt{\gamma_B/3} a_{\text{B}} m_{\text{Pl}}$ (the blue solid
curve in Fig.~\ref{length}), so we can use it to classify different modes. Some modes with large values of $k^2\gg k_\text{B}^2$ (the region below the
low (orange) dashed line in Fig.~\ref{length}) are inside the horizon all the time until they exit the Hubble horizon during the slow-roll inflation. Some of the modes
with smaller  $k^2\ll k_\text{B}^2$ (the region above the upper (green) dashed line  in Fig.~\ref{length}) exit and re-enter the horizon during the bouncing process,
and will finally re-exit the Hubble horizon during the slow-roll inflation. Since the modes with $k\gg k_\text{B}$ are inside the horizon during the whole pre-inflationary
phase, they will have the same power-law spectra as those given in GR \cite{inflation}. We are interested in the modes with $k\simeq k_\text{B}$ (the shaded region in Fig.~\ref{length}).
However, the perturbations for these modes have different behaviors when they are inside or outside the horizon, which makes Eq.~(\ref{scalar_eom}) extremely
difficult to be solved  analytically.

\begin{figure}
\includegraphics[width=8.5cm]{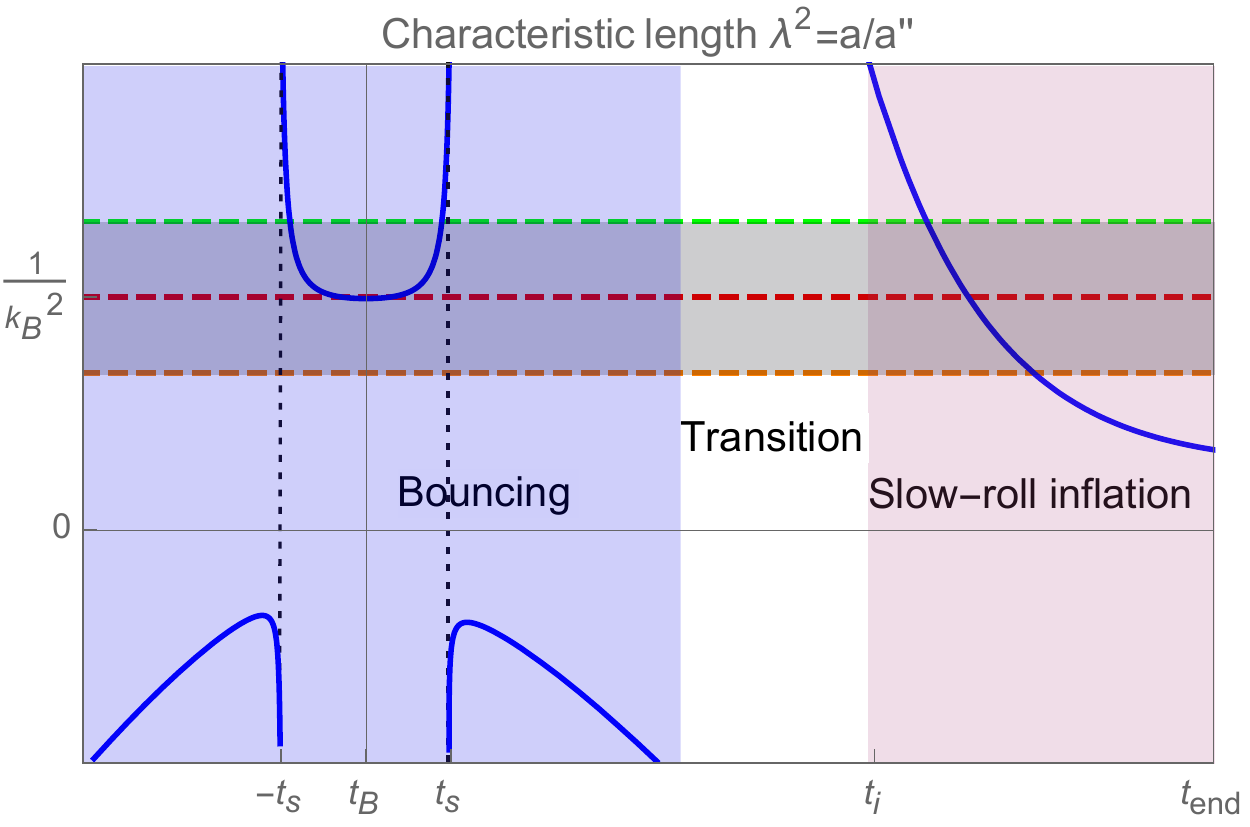}
\caption{Schematic plot of $\lambda^2 [\equiv a/a'']$, where $\left.a''/a\right|_{t=t_s}=0$ with $t_s \simeq 0.2 t_{\rm Pl}$, and $\ddot a(t_i)=0$ with $t_i$ being the starting time of the inflationary phase.
During the slow-roll inflation, $a/a'' =L_H/2$. The expansion factor  $a(t)$ can be analytically extended to a contracting phase $t<t_{\rm B}$.
} \label{length}
\end{figure}

In this Letter, we  first present an analytical solution  of  Eq.~(\ref{scalar_eom}) by using an effective P\"oschl-Teller (PT) potential. To  this goal, let us first consider the quantity,
\bqn\lb{potential}
\mathscr{V}(\eta)\equiv\frac{a''}{a}=a_\text{B}^2 \frac{\gamma_B m_{\text{Pl}}^2(3- \gamma_B t^2/{t_{\text{Pl}}^2)}}{9(1+ \gamma_B t^2 /t_{\text{Pl}}^2)^{5/3}}.
\eqn
If we consider Eq.~(\ref{scalar_eom}) as the Schr\"odinger equation, then $\mathscr{V}(\eta)$ serves as an effective barrier   during the bouncing
phase. Such a potential can be approximated by a PT potential for which we know the analytical solution,
\bqn\lb{PT}
\mathscr{V}_{\text{PT}}(\eta) = {\mathscr{V}_0}{\cosh^{-2}{\alpha (\eta-\eta_\text{B})}},
\eqn
where  $\mathscr{V}_0= a_\text{B}^2  \gamma_B m_{\text{Pl}}^2/3$ and $\alpha^2 \equiv 2 a_\text{B}^2 \gamma_B m_{\text{Pl}}^2=6 k_{\rm B}^2$.
From Fig.~\ref{PT_potential} we can see  that $\mathscr{V}_{\rm PT}(\eta)$ mimics  $\mathscr{V}(\eta)$ very well.
Introducing $x$ and $\mathcal{Y}(x)$ via   $x(\eta) = 1/(1+e^{-2 \alpha (\eta-\eta_\text{B})})$, $\mathcal{Y}(x)=[x (1-x)]^{i k/(2\alpha)} \mu_k(\eta)$, we find that Eq.~(\ref{scalar_eom}) reduces to,
\bqn\lb{Hyper_eq}
x (1-x)  \mathcal{Y}'' +[c_3-(c_1+c_2+1)x] \mathcal{Y}' -c_1 c_2 \mathcal{Y}=0, ~~~
\eqn
where $\mathcal{Y}' \equiv {d\mathcal{Y}}/{dx}$ and
\bqn\lb{a1a2a3}
&& c_1 \equiv \frac{1}{2} +\frac{1}{2\alpha} \sqrt{\alpha^2-4\mathscr{V}_0}- \frac{ik}{\alpha},\nb\\
&& c_2\equiv \frac{1}{2} - \frac{1}{2\alpha} \sqrt{\alpha^2-4\mathscr{V}_0}- \frac{ik}{\alpha},\nb\\
&& c_3 \equiv  1-\frac{ik}{\alpha}.
\eqn
This equation
is the standard hypergeometric equation, and its general solution is given by,
\bqn\lb{sol_PT}
\mu^{(\text{PT})}_k(\eta) &=& a_k x^{ik/(2\alpha)} (1-x)^{-ik/(2 \alpha)} \nb\\
&& \times \; _2F_1(c_1-c_3+1,c_2-c_3+1,2-c_3,x)\nb\\
&& +b_k [x (1-x)]^{-ik /(2 \alpha)} \;_2F_1(c_1,c_2,c_3,x). ~~~~~
\eqn
Here $a_k$ and $b_k$ are two integration constants to be determined by the initial conditions.
\begin{figure}
\includegraphics[width=8.5cm]{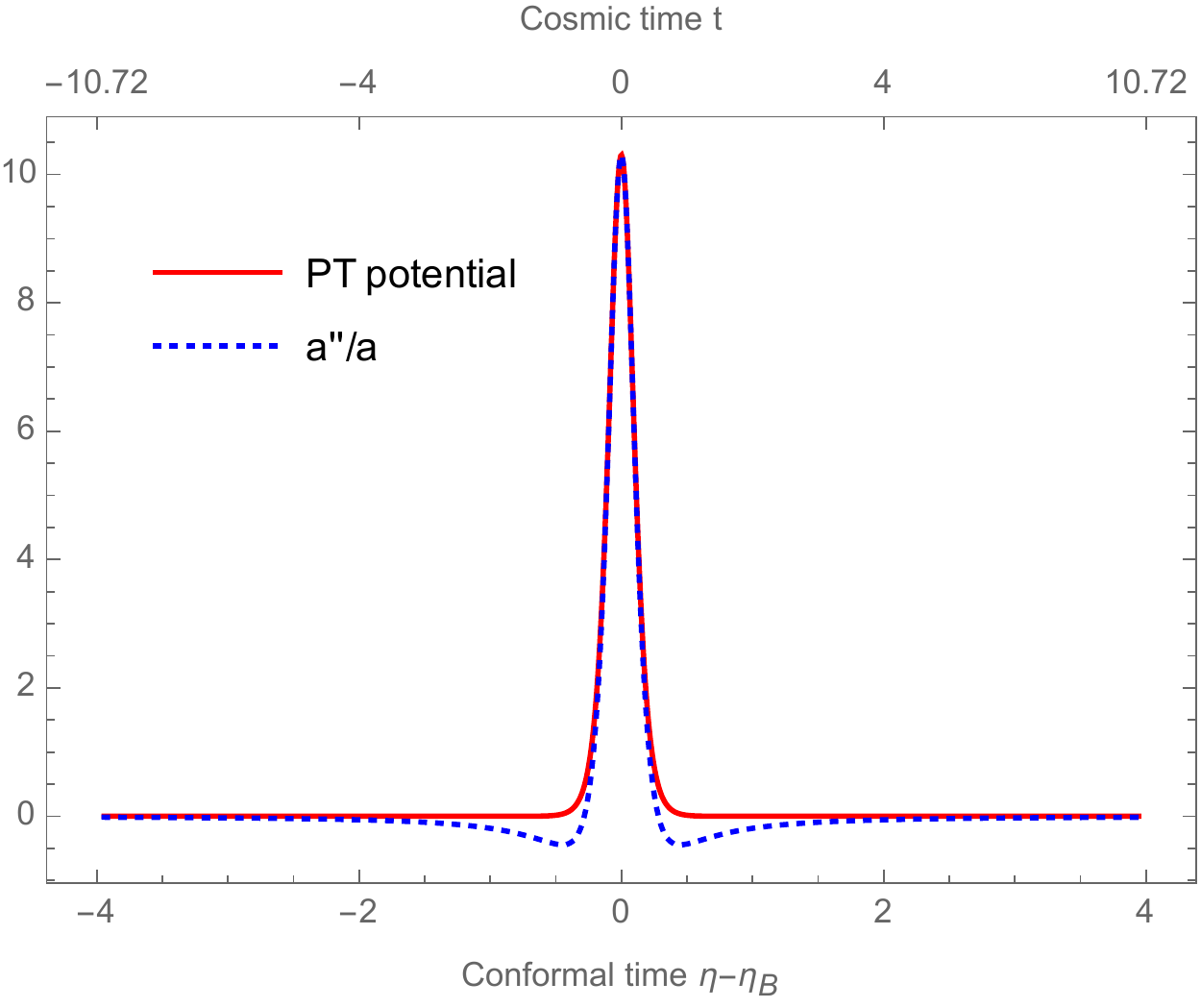}
\caption{Comparison between the effective potential given by Eq.~(\ref{potential}) and the PT potential in Eq.~(\ref{PT}). In this plot, we have set $a_{\rm B}=1, \; m_{\rm Pl}=1$.
}\label{PT_potential}
\end{figure}

\begin{figure}
\includegraphics[width=8.5cm]{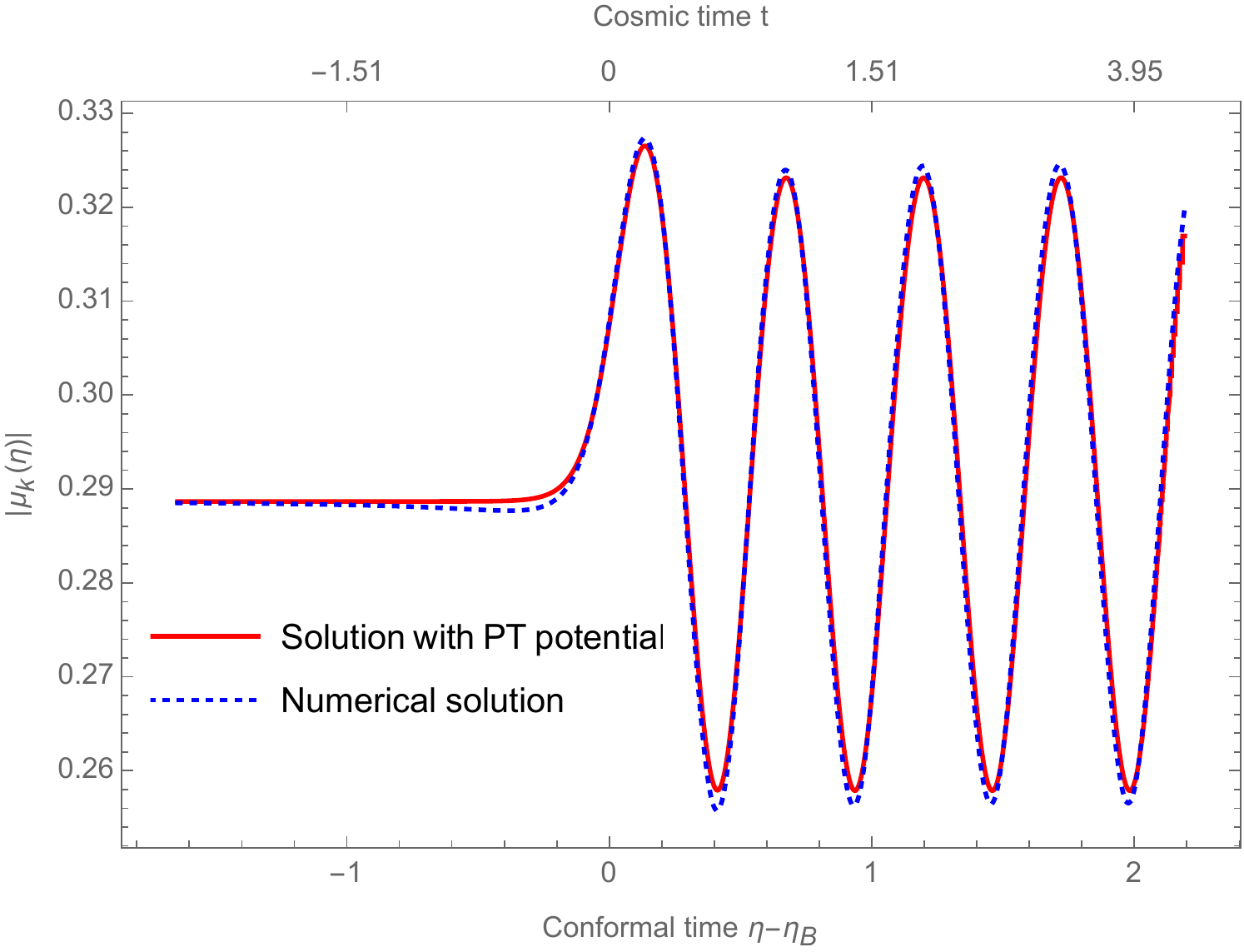}
\caption{Comparison between the analytical solution and numerical one with $a_{\rm B}=1$, $m_{\rm Pl}=1$, and  $k=6$.
}\label{compar}
\end{figure}
To impose them, let us first specify  the initial  time. A natural choice is
right at the bounce, at which  the initial state can be constructed as the {\em fourth-order adiabatic vacuum} \cite{planck_extension, planck_extension_CQG}.
 While such constructions work well for large $k$, however,  ambiguity remains
for modes with $k < k_\text{B}$ \cite{planck_extension_CQG}.
Another choice that has been frequently used is a time during the contracting phase when the modes are well within the characteristic length $\lambda$,
which is $t \lesssim -t_s$ as shown in Fig.~\ref{length}
 \cite{deformed_tensor, deformed_scalar, deformed, quadratic_loop, Starobinsky_loop,WE12}.
In this Letter, we also shall make that choice, as the main conclusions will not sensitively depend on these choices, as shown  in \cite{bounce_uniform,quadratic_loop,ANA15}, and 
we require that at this initial time the state should be the BD vacuum.
Then, we find
\bqn
\lb{CF}
a_k =0 , \; b_k = \frac{e^{i k\eta_\text{B}}}{\sqrt{2k}}.
\eqn
It should be noted that $\mu^{(\text{PT})}_k(\eta)$ of Eq.~(\ref{sol_PT}) and the above initial conditions are valid for any value of $k$. In particular,  at the bounce it
reduces to the one obtained in \cite{planck_extension_CQG}  with the {\em fourth-order adiabatic vacuum} for large $k>k_\text{B}$.  This further confirms our above arguments.
In Fig.~\ref{compar} we compare our analytical approximate solution  with the numerical (exact) one, which shows that they match extremely well
during the bouncing phase.   After this period, the universe soon sets to  the slow-roll inflation phase, and the mode functions of tensor and scalar perturbations are the
well-known solutions given  in GR \cite{inflation}. When all the relevant modes are inside the Hubble horizon
($t < t_i$ as shown in Fig.~\ref{length}), they take the asymptotic form \cite{inflation},
 \bqn\lb{sol_GR}
\mu_k^{(s,t)}(\eta) \simeq \frac{1}{\sqrt{2k}}\left(\alpha_k e^{-i k\eta}+\beta_k e^{ik\eta}\right),\; (t < t_i).
\eqn
 In GR, one usually imposes the BD vacuum at the beginning of inflation, at which  all the (physical) modes are inside the Hubble horizon, so that  $\alpha^{GR}_k = 1, \; \beta^{GR}_k = 0$. This
 in turn  leads to the standard power-law spectra.
 However, due to the quantum gravitational effects,   $\beta_k$ now does not vanish generically. To see this, we need to match   the  GR solution  to Eq.~(\ref{sol_PT}).
 Taking its limit $t/t_{Pl} \gg 1$  and then comparing it with the GR solution we find
 \bqn
 \lb{main}
 && \alpha_k=\frac{\Gamma(c_3)\Gamma(c_3-c_1-c_2)}{\Gamma(c_3-c_1)\Gamma(c_3-c_2)} e^{2 i k\eta_\text{B}},\nb\\
 && \beta_k=\frac{\Gamma(c_3)\Gamma(c_1+c_2-c_3)}{\Gamma(c_1)\Gamma(c_2)},
 \eqn
 where $c_n$ are the constants given by Eq.~(\ref{a1a2a3}).  This represents one of our main results. When $k \simeq k_B$  we find that $\left| \beta_k\right|^2 \simeq 10$.
 That is,  particles  of such modes were created during the bouncing
 phase.  However, such creation  will not alter significantly the evolution of the background, nor the perturbations during the slow-roll inflation period, as shown explicitly in
\cite{planck_extension_CQG}.
  Then, from Eq.~(\ref{sol_PT}) we obtain $\mathcal{P}^{(s, t)}_{\text{LQC}}(k)=|\alpha_k+\beta_k|^2 \mathcal{P}^{(s, t)}_{\text{GR}}(k) \equiv \left(1 + \delta_{\mathcal{P}}\right)\mathcal{P}^{(s, t)}_{\text{GR}}(k)$, where
\bqn\lb{pw}
\delta_{\mathcal{P}} &\equiv& \left[1+\cos{\left(\frac{\pi}{\sqrt{3}}\right)}\right]\text{csch}^2\left(\frac{\pi k}{\sqrt{6} k_{\rm B}}\right)\nb\\
&&+\sqrt{2}\cos{\left(\frac{\pi}{2 \sqrt{3}}\right)}\sqrt{\cos{\left(\frac{\pi}{\sqrt{3}}\right)}+\cosh\left(\frac{2\pi k}{\sqrt{6} k_{\rm B}}\right)}\nb\\
&&\times \text{csch}^2\left(\frac{\pi k}{\sqrt{6} k_{\rm B}}\right) \cos{(2k\eta_{\text{B}}+\varphi_k)},
\eqn
where 
\bqn
\varphi_k \equiv \arctan\left\{\frac{{\rm Im} [\Gamma(c_1) \Gamma(c_2) \Gamma^2(c_3-c_1-c_2)]}{{\rm Re} [\Gamma(c_1) \Gamma(c_2) \Gamma^2(c_3-c_1-c_2)]}\right\}.\nb\\
\eqn
In Fig.~\ref{spectrum}, we display the ratio between the power spectrum with { the} bounce effects and the standard power-law spectrum in GR, i.e., $1+\delta_P$ with $\delta_P$ being given by 
{the} above equation, as a function of wavenumber. We would like to {note that Fig.~\ref{spectrum} is consistent with that given in  } \cite{planck_extension, planck_extension_CQG} 
(c.f. Fig. 1 in the first paper of \cite{planck_extension} and Fig. 5 in \cite{planck_extension_CQG}). While the results obtained in \cite{planck_extension, planck_extension_CQG} are purely numerical, here ours 
are derived directly from the analytical expression {of Eq.~(\ref{pw})}. 

It is remarkable to note that, although it is well-known that quantum gravitational effects often lead to oscillations \cite{trans-planck},
{\em in LQC the  oscillating phases for both  scalar and tensor perturbations are the same}.
In Eq.~(\ref{pw}), the second term is oscillating very fast and can be ignored observationally \cite{planck_extension,planck_extension_CQG,quadratic_loop}.
On the other hand, the first term, proportional to $\text{csch}^2[\pi k/(\sqrt{6}k_{\rm B})]$, decreases exponentially as $k$ increases, and the power spectra get enhanced (reduced)  for small (large) $k$.
The modes $k\simeq k_\text{B}$,  of the Planck scale at the bounce,  are initially inside the radius defined by $\lambda=\sqrt{|a/a''|}$, and then leave and re-enter it during the bouncing phase.
The modes with $k\gg k_\text{B}$ are always inside the radius before they leave the Hubble horizon during the slow-roll inflation, thus they finally lead to a standard power spectrum.

\begin{figure}
\includegraphics[width=8.5cm]{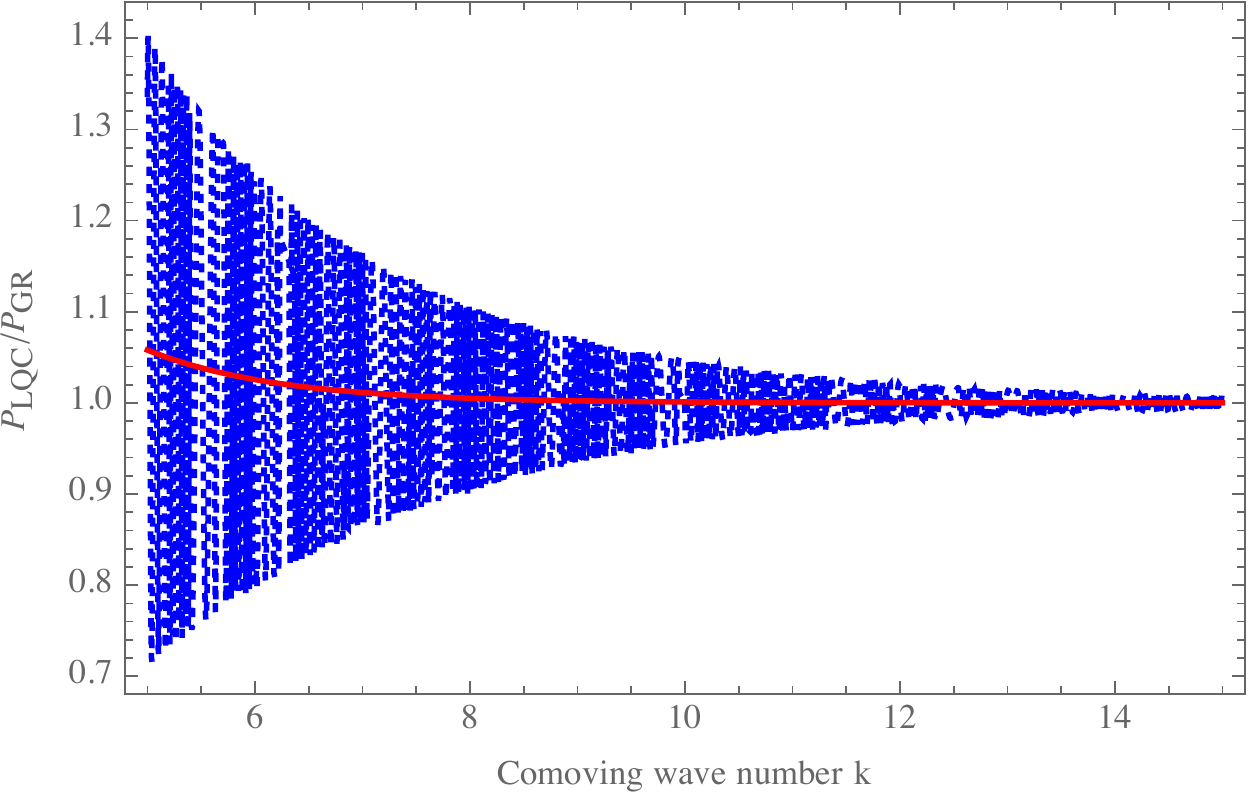}
\caption{The ratio $\mathcal{P}^{(s, t)}_{\text{LQC}}(k)/ \mathcal{P}^{(s, t)}_{\text{GR}}(k)$ between the power spectrum with the bounce effects and the standard power-law one obtained in GR. 
The dotted blue curve denotes the analytical power spectrum, which 
obviously oscillates rapidly with $k$. The solid red curve shows the average of the oscillating spectrum.
} \label{spectrum}
\end{figure}

It should also be noted that the solution with the PT potential is not valid   for the modes with a very small  $k$ (i.e., $k \ll |a''/a|$ holds all the time during the bouncing phase).
For these modes, if we ignore the $k^2$ term in Eq.~(\ref{scalar_eom}), the solution can be approximated by \cite{deformed_tensor}, 
\bq
\lb{GRr}
\mu_k(\eta) \simeq a_k a(\eta)+\frac{b_k}{a(\eta)}.
\eq 
However, we are not interested in these modes, as they currently  are still outside of the observable universe.

\begin{figure}
\includegraphics[width=8.1cm]{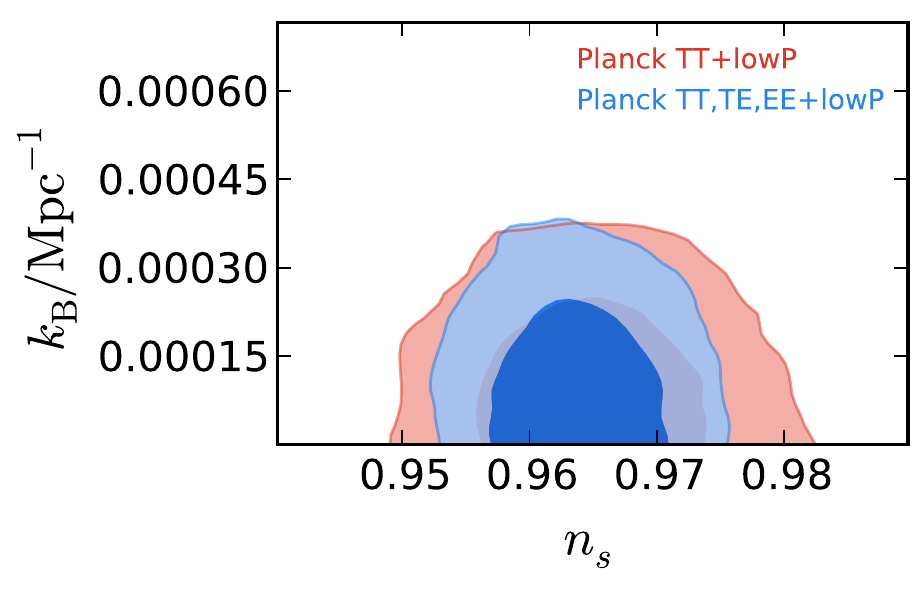}\\
\includegraphics[width=8.1cm]{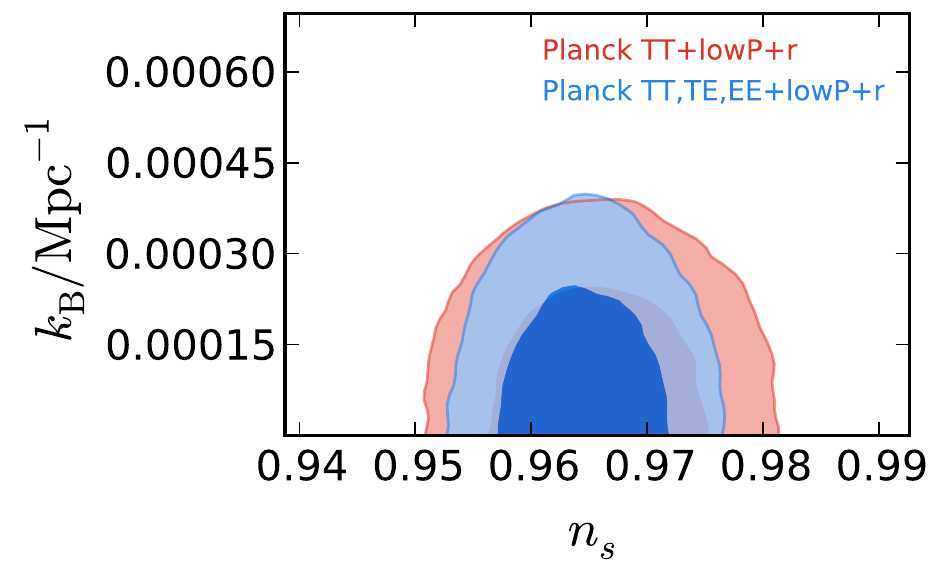}
\caption{Observational constraints for $(n_s, k_\text{B}/\text{Mpc}^{-1})$ at 68\% and 95\% C.L. by using the Planck 2015 TT+lowP and TT, TE, EE+lowP data with $a_0=1$.
The upper panel only considers the scalar spectrum, while the bottom  includes  the tensor.
} \label{CMB_Likelihood}
\end{figure}

 \section{ Observational Constraints}
 
  \begin{table*}
\caption{The Best fitting values of the six cosmological parameters and the constraints on $k_\text{B}/a_0$ and $r$ at 95\% C.L for different cosmological models from different data combinations.}
\lb{bestfit}
\begin{ruledtabular}
\begin{tabular}{ccccc}
 Parameter  & Planck TT+lowP & Planck TT,TE,EE+lowP & Planck TT+lowP+$r$ & Planck TT,TE,EE+lowP+$r$ \\
\hline
$\Omega_{\rm b}h^2$ &  $0.022355$ & $ 0.022193$ & $0.022322 $ & $0.022064$\\
$\Omega_c h^2 $  & $0.11893$ & $ 0.12000 $ & $0.11908$ & $0.12071 $\\
$100\theta_{\rm MC}$& $1.04115$ & $ 1.04065 $ & $ 1.04080$ & $1.04057$ \\
$\tau$ &  $0.077835$ & $0.089272 $ & $0.081955$ & $0.085259$   \\
$\ln(10^{10}A_s)$ &  $3.088$ & $3.112 $ & $3.101$ & $3.104 $ \\
$n_s$ &  $0.9662$ & $ 0.9647 $ & $ 0.9658$ & $ 0.9607 $ \\
\hline
$k_{\rm B}/a_0 $ & $<3.12\times 10^{-4}$& $ <3.05\times 10^{-4}$& $<3.14\times 10^{-4}$& $< 3.14 \times 10^{-4}$ \\
$ r$ &  $ ---- $& $----$ & $ < 0.113 $& $ < 0.107 $\\
\end{tabular}
\end{ruledtabular}
\end{table*}

The quantum corrections (\ref{pw})  are $k$-dependent and expected to be constrained  by observations. In the following, we perform the CMB likelihood analysis by using the Planck 2015 data \cite{Planck2015},
 with the MCMC code developed in \cite{cosmomc}.
 We assume the flat cold dark matter model with the effective number of neutrinos $N_{\text{eff}}=3.046$ and choose the total neutrino mass as $\Sigma m_\nu=0.06{\text{eV}}$. We also write
 \bq
 \lb{PP}
 \mathcal{P}_{\text{GR}}^{(s,t)}(k)=A_{(s,t)} \left(\frac{k}{k_*}\right)^{n^{(s,t)}_{\text{inf}}}, 
 \eq
 where $k_*(=0.05 {\rm Mpc}^{-1})$ denotes the pivot scale,  $n^{(s)}_\text{inf}=n_s-1$ and $n^{(t)}_\text{inf}=n_t$.
 We vary the seven parameters, $\Omega_{\rm b}h^2, \Omega_\text{c}h^2, \tau, \Theta_s, n_s, A_s, k_\text{B}/a_0$ \cite{Zhu2}. 
 For the six cosmological parameters except $k_{\rm B}/a_0$ ($\Omega_{\rm b}h^2, \Omega_\text{c}h^2, \tau, \Theta_s, n_s, A_s)$, we use the same prior ranges as those adopted
  in \cite{Planck_parameters}, while for the parameter $k_{\rm B}$ which is related to the bouncing effects, we set the prior range to $k_{\rm B} \in [10^{-8}, 0.002]$.
 
 In particular, we use {the} high-$l$ CMB temperature power spectrum (TT) and polarization data (TT, TE, EE) respectively with {the} low-$l$ polarization data (lowP) from Planck2015.
  In Table. \ref{bestfit}, we list the best fit values of {the} six cosmological parameters and constraints on $k_{\rm B}/a_0$ and $r$ at $95\%$ C.L. for different cosmological models from different data combinations. 
 
Marginalizing other parameters, we find that $k_\text{B}/a_0$ is constrained by the Planck TT+lowP (Planck TT,TE,EE+lowP) to 
\bq
\lb{kba0}
\frac{k_\text{B}}{a_0} < 3.12 \times 10^{-4}\text{Mpc}^{-1} (3.05 \times 10^{-4}),
\eq
 at 95\% C.L
[cf. Fig.~\ref{CMB_Likelihood}]. When we consider the ratio $r=A_{(t)}/A_{(s)}$, the Planck TT+lowP (Planck TT,TE,EE+lowP) data yields 

\bq
\lb{kBk02}
\frac{k_{\text{B}}}{a_0}
< 3.14 \times 10^{-4}\text{Mpc}^{-1} (3.14 \times 10^{-4}),
\eq
 at 95\% C.L. These upper bounds show that the observational constraints on the bouncing effects are robust with respect 
 to different data sets (without/with the polarization data included) and whether the tensor spectrum is included or not. In Fig.~\ref{triA} we show constraints on  a couple of cosmological parameters and their respective probability 
 distributions for the CosmoMC runs described above and for the results from {the} Planck 2015 data. We notice that the colored curves which represent the probability distributions of $k_{\rm B}/a_0$ are almost 
 perfectly superposed, which strongly indicates again that the constraints on $k_{\rm B}$ derived in this paper are robust. 

Using {the} relation
\bqn
\frac{k_\text{B}}{a_0} = \sqrt{\frac{\gamma_{\rm B}}{3}} \frac{a_{\rm B}}{a_0} m_{\rm Pl}=\sqrt{\frac{\gamma_{\rm B}}{3}} m_{\rm Pl} e^{-N_{\rm tot}},
\eqn
where $N_{\rm tot} \equiv \ln{(a_0/a_{\rm B})}$ denotes the total e-folds from the quantum bounce until today, then {the} above upper bounds on $k_{\rm B}/a_0$ can be translated into the constraint on the  total $e$-folds 
$N_{\text{tot}} $ as
\bqn
\lb{NT}
N_{\rm tot}>141  \;\; (95\% {\rm C.L.}), 
\eqn
where we have taken
$\rho_\text{c}=0.41 m_\text{Pl}^4$ \cite{planck_extension,planck_extension_CQG}. This in turn leads to a lower bound of $\delta N_*$, 
\bq
\delta N_* > N_{\text{tot}}-N_*-N_\text{after}, 
\eq
where $\delta N_* \equiv \ln{(a_*/a_B)}$, $N_* \equiv \ln{(a_{\text{end}}/a_*)}$, and $N_\text{after} \equiv \ln{(a_0/a_{\text{end}})}$, where  $a_*$ denotes  the expansion factor at the moment that the current Horizon exited the Hubble horizon during the slow-roll inflation, and $a_{\text{end}}$ is that of the end of inflation. Taking $N_*\simeq 60\simeq N_\text{after}$, we find  
\bq
\lb{NP}
\delta N_* \gtrsim 21.
\eq

Note that our results given by Eqs.(\ref{NT}) and (\ref{NP}) are based on  three assumptions: (1) the Universe is filled with a scalar field with its potential $V(\phi)$; (2)  the background evolution
initially  is dominated completely  by the kinetic energy of the scalar field; and  (3)  the Universe is  in the BD vacuum  state  in the contracting phase ($t \lesssim - t_s$, as shown in Fig.~\ref{length}).

\begin{figure*}
\includegraphics[width=14.5cm]{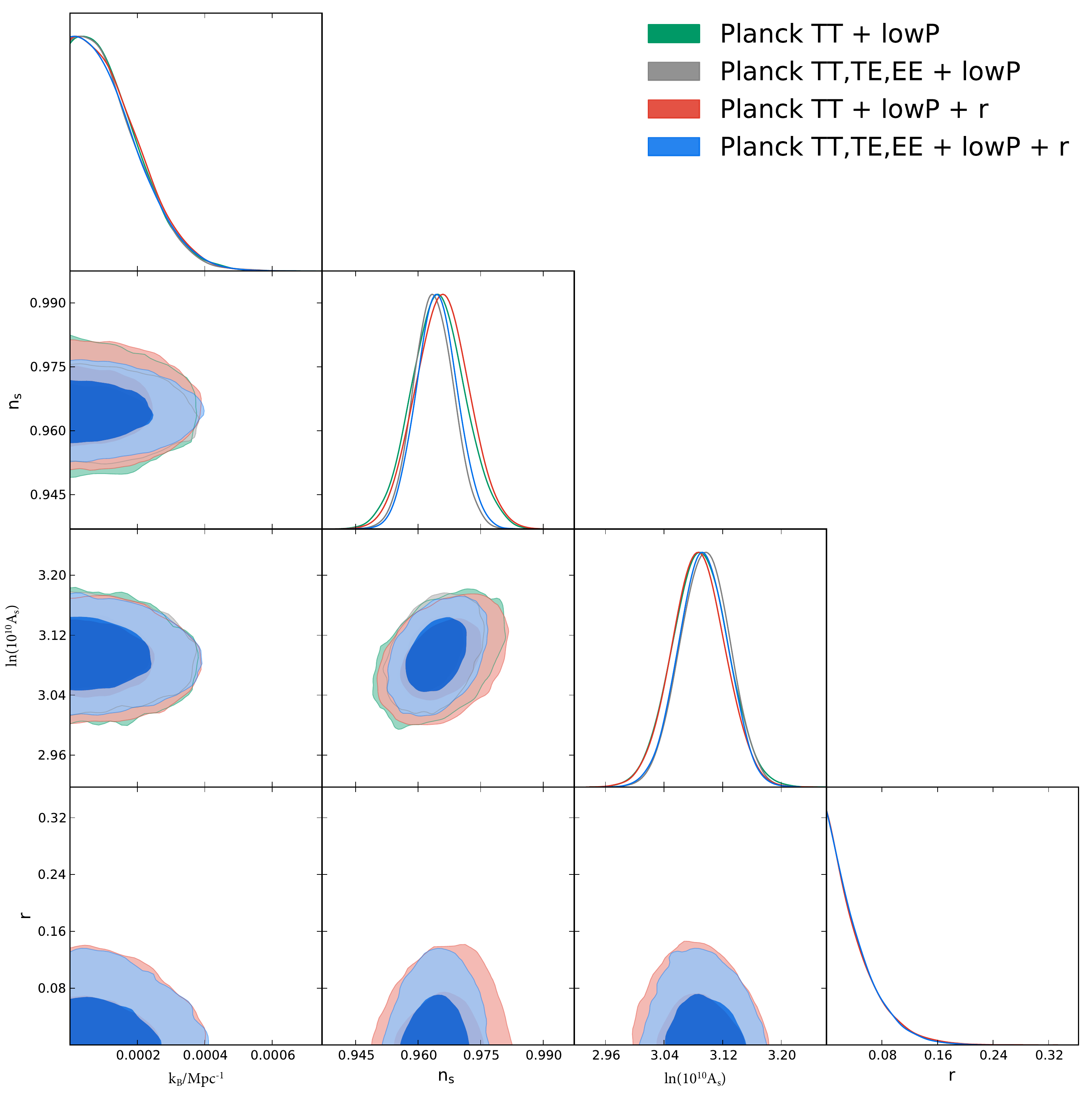}
\caption{Observational constraints on  {a couple of} parameters ($68\%$ and $95\%$ contour lines) and { the} probability distributions for $\ln(10^{10} A_s)$, $n_s$, $k_{\rm B}/a_0$, and $r$ by 
using the Planck 2015 data. Note that in {the} numerical simulations we set $a_0=1$.
} \label{triA}
\end{figure*}

 \section{ Conclusions}

 In this Letter, we  {\em analytically} studied the evolutions of the background and the linear scalar and tensor perturbations of the FLRW universe in LQC within the framework of {\em the  dressed metric approach} \citep{planck_extension,planck_extension_CQG,quadratic_loop}, and showed that,   { if the pre-inflationary phase  is dominated by the kinematic energy of the inflaton, the evolutions 
 will be    {\em  independent of  the slow-roll inflationary models} during this phase [cf. Fig.~\ref{scalar_factor} and Eqs.~(\ref{Bsolution}) and (\ref{pw})]. Imposing the BD vacuum in the contracting phase ($t \lesssim -t_s$ as shown in
 Fig. 2), } we obtained the Bogoliubov coefficients (\ref{main}) at the onset of the slow-roll inflation, which shows clearly that   during the pre-inflationary phase, particles are generically  created ($\left.\beta_k\right|_{t=t_i} \not= 0$),
 and the resulting power spectra are $k$-dependent. This is in contrast to GR (where the BD vacuum ($\left.\beta_k\right|_{t=t_i} = 0$) is  usually imposed at the onset of the slow-roll inflation
 \cite{inflation}.  {This provides a potential window to   test LQC  directly by the measurements of CMB and galaxy surveys \citep{S4-CMB}. In particular,} fitting the power spectra to   the Planck 2015 temperature (TT+lowP)
 and polarization (TT,TE,EE+lowP) data, we found the lower bound for $N_{\text{tot}} \equiv \ln{(a_0/a_B)} >141$ (95\% C.L.). That is, to be consistent with current observations of CMB,  the universe must have expanded
 at least 132 e-folds since the bounce.

\section*{Acknowledgements}

We would like to thank M. Sasaki and W. Zhao for valuable comments and suggestions. This work is supported in part by Ciencia Sem Fronteiras, Grant No. 004/2013 - DRI/CAPES, Brazil (A.W.); Chinese NSF Grants, Nos. 11375153 (A.W.),
11675145(A.W.), 11675143 (T.Z.), 11105120 (T.Z.), and 11205133 (T.Z.).

\end{document}